\documentstyle[preprint,eqsecnum,aps]{revtex}

\def\ehti{\raisebox{.7mm}{\hbox{\tiny 1}}
        \hbox{\tiny /}
        \raisebox{-.7mm}{\hbox{\tiny 2}}}

\def\ehss{\raisebox{.7mm}{\hbox{\scriptsize 1}}
        \hbox{\scriptsize /}
        \raisebox{-.7mm}{\hbox{\scriptsize 2}}}

\newcommand{\be}{\begin{eqnarray}}
\newcommand{\ee}{\end{eqnarray}}

\newcommand{\cP}{{\cal P}}
\newcommand{\cD}{{\cal D}}

\newcommand{\cM}{{\cal M}}
\newcommand{\cH}{{\cal H}}
\newcommand{\bi}{\bigskip}
\newcommand{\no}{\noindent}
\newcommand{\hk}{\hspace{0.1cm}}
\newcommand{\hs}{\hspace{0.5cm}}
\newcommand{\dslash}{\partial\hskip -0.5em/}

\newcommand{\rk}{\right)}
\newcommand{\lk}{\left(}

\newcommand{\sli}{\sum\limits}
\newcommand{\pli}{\prod\limits}
\newcommand{\il}{\int\limits}

\begin{document}
\draft
\date{\today}
\preprint{{UNITU-THEP-09/1996}}
\title{Yang-Mills Theory In Axial Gauge }
\author{H. Reinhardt} 
\address{Institut f\"ur Theoretische Physik, Universit\"at T\"ubingen\\
Auf der Morgenstelle 14, D-72076 T\"ubingen, Germany\footnote{permanent
address} \\
and\\
Center for Theoretical Physics\\
Laboratory for Nuclear Science and Department of Physics\\
Massachusetts Institute of Technology\\
Cambridge, Massachusetts 02139, USA}
\maketitle

\begin{abstract}
The Yang-Mills functional integral is studied in an axial variant of 't
Hooft's maximal Abelian gauge. In this gauge Gau\ss ' law can be
completely resolved resulting in a description in terms of unconstrained
variables. Compared to previous work along this line starting with work
of Goldstone and Jackiw one ends up here with half as many integration
variables, besides a field living in the Cartan subgroup of the gauge
group and in D-1 dimension. The latter is of particular relevance for
the infrared behaviour of the theory. Keeping only this variable we
calculate the Wilson loop and find an area law.
\end{abstract}

\section{Introduction}
\bi

\noindent
It is a common belief, that the fundamental interactions are
described by gauge theories. This is, in particular, true for strong
interactions, which are assumed to be described by QCD. This theory has
been tested in the high energy regime, where perturbation theory is
applicable due to asymptotic freedom. On the other hand, low energy
hadron physics requires a non-perturbative treatment of QCD. This
regime, which is alternatively related to the confinement problem, is
much less understood. Perturbative calculations indicate, that the
confinement phenomenon is due to the non-abelian nature of Yang-Mills
theory. Furthermore, one-loop calculations show, that the perturbative
Yang-Mills vacuum is unstable \cite{sav} and various mo\-dels of the
Yang-Mills vacuum have been designed as, for example, the Copenhagen
vacuum \cite{niels} the instanton liquid model \cite{shur}, which extends the
instanton gas picture \cite{dashen}, the dual superconductor \cite{super}
or the
stochastic vacuum \cite{dosch}. The various models aim at different
aspects of strong interaction, e.g. the instanton models seem to be
suited to explain spontaneous breaking of chiral symmetry
\cite{leutwyler}, while the Copenhagen vacuum, stochastic vacuum or the
dual superconductor focus on the color confinement.
\bi

\no
A rigorous approach to strong coupling Yang-Mills theory is provided by
lattice Monte-Carlo calculations \cite{lattice}, which 
have been developed to a high
level of sophistication. This approach has given much insight into the
nature of the Yang-Mills vacuum. The great successes of lattice
calculations are in low-energy hadron physics (where confinement is
perhaps not of much relevance) \cite{lattice94}.
\bi

\no
There has been much progress in controlling finite size effects. This
concerns both the fermion description on lattice as well as improved
lattice actions for gauge fields \cite{breiten}. 
However, a complete understanding of the
Yang-Mills theory will probably not be provided by the lattice
simulations alone but requires also analytic tools. For some
applications of lattice QCD a separation of scales is required and input
from perturbation theory is needed (see e.g. \cite{lat-struct}).
Also the interpretation of the lattice results sometimes requires or at
least is facilitated by modelling properties of the Yang-Mills-vacuum
like correlation functions, which in turn are fed by the lattice
calculations (see e.g. \cite{giacomo}).
\bi

\no
Several analytic approaches have been proposed to explore the
non-perturbative features of strong coupling
Yang-Mills theory. In particular, there is the strong coupling 
expansion 
itself for lattice QCD \cite{lattice}. In the strong coupling limit one 
finds Wilson's 
area law (or a linear rising quark potential), which is considered as a
signal of
confinement. However, the strong coupling limit yields the area law for
almost any gauge group and in any dimension, but the finite convergence
radius of the strong coupling expansion forbids the strong coupling
result to be applied to the continuum limit. In fact the finite
convergence radius is taken as an explanation to argue away
confinement in cases where the property of confinement is
undesirable, as e.g. in QED.
\smallskip

\no
L\"uscher at al. \cite{lusch} have proposed a small volume expansion
exploiting the fact, that perturbation theory becomes eventually 
applicable as the volume decreases \cite{koller}.
\bi

\no
Most analytic approaches to Yang-Mills theory are based on the Weyl
gauge $A_0 = 0$, where the Gau\ss '  law has to be inforced as a
constraint to guarantee local gauge invariance \cite{jackiw}. Violation 
of Gau\ss ' law
generates color charges during the time evolution and, by this leaking of
color, confinement is lost. This fact has recently been emphasized in
reference \cite{johnson1}, where explicit projection on gauge invariant
states has been performed in the construction of the path integral
representation of the Yang-Mills transition amplitude. Not surprisingly
projection onto gauge invariant states requires a compact integration
measure (over the Haar measure of the gauge group), reminiscent to the
lattice approach. In fact, the approach of ref. \cite{johnson1} can be obtained
from the lattice formulation by taking the continuum limit in the
spatial directions only. 
\bi

\no
Several approaches have
been advocated, which explicitly resolved the Gau\ss ' law constraint by
changing variables resul\-ting in a description in terms of a reduced number of
unconstraint variables. These approaches are based on the Schr\"odinger
functional formulation of Yang-Mills theory \cite{jackiw}.
Refs. \cite{goldstone}, \cite{faddeev}
use variants of unitary
gauge, while in ref. \cite{lee} the Coulomb gauge $\nabla A = 0$ was used.
Refs. \cite{goldstone}, \cite{faddeev} 
basically end up in a description in terms
of gauge invariant variables (Further approaches along these lines are
proposed in \cite{simonov},
\cite{badalyan}.
\bi

\no 
Recently an alternative descriptions of Yang-Mills
theory in terms of gauge invariant variables constructed either from
the magnetic \cite{johnson2} or electric \cite{bauer} 
fields have been proposed. 
In ref. \cite{vauth}
the long wave length (strong
coupling) limit of the formulation of ref. \cite{johnson2}, has been studied,
exploiting methods from the description of collective excitations of
atomic nuclei. 
\bi

\no
A crucial point in all analytical approaches to Yang-Mills theory 
is the gauge fixing, which
in non-abelian gauge theories cannot be performed in a unique way due to the
existence of Gribov copies \cite{gribov}. 't Hooft has proposed a
so-called maximum abelian gauge \cite{thooft}, which partly eliminates 
the non-abelian
components of the gauge field (abelian projection). In this gauge
monopoles occur and it is believed, that their condensation yields
confinement via the dual Meissner effect, see also \cite{mandel}. This
approach has been further elaborated in \cite{schier}, where the first
lattice calculations along this line have been performed. In fact, 
recent lattice
calculations performed in the maximum abelian gauge show evidence for
monopole condensation \cite{mon}. 
\bi

\no
Recently, QCD on a spatial torus has been considered in the Weyl 
gauge $A_0 = 0$
in the cano\-nical quantization approach \cite{lenz1}. Using a (partial) 
axial gauge,
which, in fact, is a variant of a maximum abelian gauge, the resolution of
Gau\ss ' law has been achieved by applying unitary gauge transformations,
which rely on the quantum field operators. This results in a Schr\"odinger
description
in terms of unconstraint variables, where the resulting Hamiltonian is
non-local. It is fair to say, that at the moment we have little
experience with solving functional Schr\"odinger equations in quantum
field theories. Furthermore, even in the functional (operator) approach
matrix elements are given by D-1-dimensional functional integrals.
Therefore it might be more convenient to use the (D-dimensional)
functional integral representation from the very beginning.
In the
present paper I perform a resolution of 
Gauss' law analogous to ref. \cite{lenz1} but in the 
functional integral representation of Yang-Mills theory. It is the hope, 
that the
functional integral formulation will facilitate in finding 
appropriate approximation schemes. Furthermore, the functional integral
approach provides more direct excess to the topological
properties of the Yang-Mills vacuum and to numerical simultations,
exploiting Monte Carlo techniques.
\bi

\no
The balance of the paper is as follows: In order to set notation and
conventions, in section 2 the functional integral description of
Yang-Mills theory is briefly reviewed and some relevant features are
discussed.
In sect. 3 and 4 we fix the gauge and resolve the Gau\ss ' law
constraint. The Faddeev-Popov determinant is evaluated in sect. 5. In
sect. 6 the Wilson loop is evaluated thereby including only the dominant
infrared unconstrained degrees of freedom. In sect. 7 the electric field
variables are integrated out resulting in a theory in unconstrained 
degrees of freedom of the gauge potential. A short summary and some
concluding remarks are given in sect. 8. Some calculations are relegated
to Appendices.
\bi

\section{Hamiltonian formulation of gauge theories}
\bi

We consider the  gauge group $G=SU (N)$ with anti-hermitian generators $T^a$
satisfying the commutation relation
\be
\label{G1}
\left[ T^a, T^b \right] = f^{abc} T^c \hk ,
\ee
where 
$f^{abc}$ are the structure constants. We choose the standard 
normalization
\be
tr \hk \lk T^a T^b \rk = - \frac{1}{2} \delta^{ab} \hk .
\ee
Later on we will also make use of the generators in the adjoint
representation defined by 
\be
\lk \hat{T}^a \rk_{bc} = - f^{abc} \hk ,
\ee
which satisfy the same commutation relation (\ref{G1}). Throughout the
paper we shall indicate the adjoint representation by the  hat ,,  \^
\hk ''. For a quantity $\chi^a$ living in the gauge group we define the
fundamental and adjoint representation, respectively, by 
\be
\chi = \chi^a T^a \hs , \hs \hat{\chi} = \chi^a \hat{T}^a \hk .
\ee
We also use the Cartan decomposition of the gauge group
\be
G = H \otimes G / H \hk ,
\ee
where $H = U (1)^{N-1}$ denotes the Cartan subgroup of the 
gauge group and $G / H$ is
the corresponding coset space. Furthermore, we use the indices with
index zero, $a_0, b_0, \dots$ to denote a generator of the Cartan
subgroup $T^{a_0} \in H \hs , \hs \left[ T^{a_0}, T^{b_0} \right] = 0$,
while the indices $\bar{a}, \bar{b}, \dots$ are reserved for generators
of the coset space, $G/H$.
Accordingly the gauge potential $A_i (x)$ is decomposed as
\be
A_i = A^{n}_{i} + A^{ch}_{i} \hk , 
\ee
where
$ A^{n}_{i} = A^{a_0}_{i} T^{a_0}$ is the gauge potential of the Cartan
subgroup $H$ and $ A^{ch}_{i}$ lives in the coset space
$G/H$. With respect to (color) charges of the Cartan subgroup $H$, $
A^{n}_{i}$ is neutral, while $ A^{ch}_{i}$ is charged.
\bi

\no
We also introduce the covariant derivative by
\be
D_\mu = \partial_\mu + A_\mu
\ee
and the field strength tensor
\be
F_{\mu \nu} & = & \left[ D_\mu, D_\nu \right] = F^{a}_{\mu \nu} T^a
\nonumber\\
 F^{a}_{\mu \nu}  & = & \partial_\mu A^{a}_{\nu} - \partial_\nu
A^{a}_{\mu} + f^{abc} A^{b}_{\mu} A^{c}_{\nu} \hk .
\ee
Under a gauge transformation $\Omega (x) \in SU (N)$ the gauge
potential transforms as
\be
A_\mu \longrightarrow A^{\Omega}_{\mu} = \Omega \lk D_\mu \Omega^\dagger
\rk = \Omega A_\mu \Omega^\dagger + \Omega \lk \partial_\mu
\Omega^\dagger \rk \hk .
\ee
In the Hamilton formulation of Yang-Mills theory, which is based on the
Weyl gauge
\be
A_0 (x) = 0 \hk ,
\ee
the dynamical variables
are the spatial components of the gauge potential $A^{a}_{i}$.
We shall use spatially periodic boundary conditions for the field
variables 
\be
\label{12}
A^{a}_{i} \lk x + L_k {\bf e}^k \rk = A_i (x) 
\ee
where ${\bf e}^k$ denotes 3-dimensional (spatial) unit vector, so that we 
consider
Yang-Mills theory on a 3-dimensional torus. We have not yet specified
the boundary condition in time direction. 
\bi

\no
Let $| C \rangle $ denote an
eigenstate of $A_i (x)$, i.e. $A_i (x) | C \rangle = C_i (x) | C \rangle
$ where $C_i (x)$ is a classical field function.
The gauge invariant transition amplitude between static initial and final field
configurations $ A_i \lk x_0 = 0, \vec{x} \rk = C'_i \lk \vec{x} \rk$ and  
$ A_i \lk x_0 = T, \vec{x} \rk = C''_i \lk \vec{x} \rk$ is
defined by  \cite{rossi}, \cite{johnson1}, \cite{luscher}
\be
Z \left[ C'', C' \right] & = & \langle C'' | e^{- HT} P | C' \rangle
\ee
where
\be
H = \int d^3 x \lk \frac{g^2}{2} E^{a}_{i} (x) E^{a}_{i} (x)
 + \frac{1}{2 g^2} B^{a}_{i} (x)   B^{a}_{i} (x) \rk
\ee
is the Yang-Mills Hamiltonian with bare coupling constant $g$, electric
field $E^{a}_{k} (x) = \delta / i \delta A^{a}_{k} (x)$ and magnetic field
$B^{a}_{k} (x) = \frac{1}{2} \epsilon_{k i j} F^{a}_{i j} (x)$.
Furthermore $P$ 
is the projector onto gauge invariant states
\be
\label{14}
P | C \rangle = \sli_n e^{- i n \Theta} \il_G \cD \mu \lk \Omega_\mu
\rk | C^{\Omega_\mu} \rangle  
\ee
Here $\Theta$ is the
vacuum angle \cite{jackiw} and the functional integration with respect to the
Haar measure $\mu (\Omega)$ of the gauge group extends over all
time-independent gauge transformation $\Omega_n \lk \vec{x} \rk$ with
winding number $n$. For a gauge transformation $\Omega \lk \vec{x} \rk$
the winding number is
defined by
\be
n [\Omega] = \frac{1}{24 \pi^2} \int d^3 x \epsilon_{ijk} tr \lk R_i R_j
R_k \rk \hs , \hs R_k = \Omega \partial_k \Omega^\dagger \hk .
\ee

\bi
\no
As usual we assume here that the gauge function $\Omega \lk \vec{x} \rk$
approaches a unique value $\Omega_\infty$ for $|\vec{x} | \to \infty$,
so that $R^3$ can be compactified to $S^3$ and $n [\Omega]$ is a
topological invariant.
\bi

\no
For many purposes it is sufficient to consider the partition function
\be
\label{C3}
Z = \int \cD C_i \langle C | e^{- HT} P | C \rangle \hk , 
\ee
which can be easily reduced to the standard form
\be
\label{C1}
Z = \sli_k e^{- E_k T} 
\ee
with $E_k$ being the energy eigenvalues. Using the completeness
of the eigenstates $| k \rangle$ of $H \lk H | k \rangle = E_k | k
\rangle \rk$ and $P^2 = P$ it can be rewritten as
\be
\label{C2}
Z = \int \cD C_i \sli_k \Psi_k (C) e^{- E_k T} \Psi^{*}_{k} (C) \hk ,
\ee
where
\be
\Psi_k (C) = \langle C | P | k \rangle
\ee
are the gauge ``invariant'' energy eigenfunctionals, which under a gauge
transformation $\Omega_n$ with winding number $n$ transform as
\be
\Psi_k \lk C^{\Omega_n} \rk = e^{- i n \Theta} \Psi_k (C) \hk ,
\ee
as is easily inferred from the explicit form of the projector (\ref{14}).
Assuming proper normalization of the energy eigenfunctionals $\Psi_k
(C)$, i.e.
\be 
\int \cD C_i \Psi^{*}_{k} (C) \Psi_l (C) = \delta_{kl}
\ee
eq. (\ref{C2}) reduces to the standard form (\ref{C1}).
\bi

\no
In ref. \cite{rein} it was explicitly shown that the gauge invariant partition
function (\ref{C3}) is given by the standard functional
integral representation
\be
\label{C4}
Z = \int \cD A_\mu \delta_{gf} e^{- S_{YM} [A] + i \Theta \nu [S]} \hk ,
\ee
where
\be
S_{YM} [A] = \frac{1}{4 g^2} \int d^4 x F^{a}_{\mu \nu} (x) F^{a}_{\mu
\nu} (x)
\ee
is the usual Yang-Mills action and
\be
\nu [A] = \frac{1}{32 \pi^2} \int d^4 x  F^{a}_{\mu \nu}   F^{a^*}_{\mu
\nu} 
\ee
is the Pontryagin index with $F^{*}_{\mu \nu} = \frac{1}{2}
\epsilon_{\mu \nu \kappa \lambda} F_{\kappa \lambda}$ being the dual
field strength. The functional integration runs over all temporally
periodic gauge field configurations $A_\mu \lk x_0 = T \rk = A_\mu \lk
x_0 = 0 \rk$ and it is understood that the gauge fixing is included by
the Faddeev-Popov method as indicated in (\ref{C4}) by $\delta_{gf}$. 
\bi

\no
At first sight one may wonder that eq. (\ref{C4}) reproduces the
gauge invariant partition function (\ref{C3}) inspite of the missing
Haar measure\footnote{In ref. \cite{johnson1} it was claimed that the
conventional functional integral representation (\ref{C4}) falls short
of guaranteeing gauge invariance in the non-perturbative regime.}.
However, as explicitly shown in \cite{rein} the Haar 
measure arises
from the Faddeev-Popov determinant. Similar investigations have been
previously performed in ref. \cite{rossi}.
\bi

\no
The equivalence proof between eqs.  (\ref{C3}) and  (\ref{C4})
\cite{rein} relies only on the gauge invariance of the Hamiltonian and
holds therefore also true when fermions are included. In this case the
partition function is given by
\be
Z =  \int \cD q \cD \bar{q} \int \cD A_\mu e^{ \int \bar{q} i\dslash q -
S_{YM} [A] + i \Theta \nu [A]} \hk ,
\ee
where the 
fermion fields satisfy antiperiodic boundary conditions
$q \lk x_0 = T \rk ~ =$ \\ $- q ( x_0 = 0 )$.
For later convenience we rewrite the partition function as
\be
Z = \int \cD q \cD \bar{q} e^{\int \bar{q} i \dslash q} Z_{YM} [J] \hk ,
\ee
where
\be
\label{C5}
Z_{YM} [J] = \int \cD A_\mu e^{- S_{YM} [A] + \int J_\mu A_\mu + i
\Theta \nu [A]} 
\ee
is formally the partition function of a gauge field coupled to an
external color current $J^{a}_{\mu} \equiv \bar{q} \frac{\lambda^q}{2}
\gamma_\mu q$. Eq. (\ref{C5}) defines  the Lagrange representation,
which is fully covariant. For subsequent considerations it is more
convenient to use the Hamilton formulation which arises from  (\ref{C5})
by linearizing the $\lk F_{oi}  \rk^2$ term by means of an integration over
the electric field variable $E^{a}_{i} (x)$ which in view of eq.
(\ref{12}) has to satisfy spatially periodic boundary condition $E_i \lk
x + L e_i \rk = E_i (x)$. Then the $A_0$ field can be
integrated out yielding the Gau\ss ' law constraint
\be
\int \cD A_0 \exp i \int d^4 x A^{a}_{0} (x) \Gamma^a (x) = \delta \lk
\Gamma^a \rk \hk ,
\ee
where
\be
\label{30}
\Gamma (A, E) = \partial_i E_i + \left[ A_i, E_i \right] + J_0 \equiv
\left[ D_i, E_i \right] + J_0 
\ee
is the generator of infinitesimal gauge transformations. Eq.  (\ref{C5})
then becomes the Hamilton functional integral representation of the
partition function
of Yang-Mills theory in the presence of an external source $J^\mu = \lk
J^0, J^i \rk$, which after continuing to Minkowski space and assuming
$\Theta = 0$ reads\footnote{In fact in the derivation of the
path integral representation  (\ref{C5}) the Hamilton formulation
(\ref{G5}) arises in an intermediate step of the calculation. In the
present case the Hamilton and Lagrange form are obviously completely
equivalent. But in more general cases (e.g. in theories with momentum
dependent masses) the Hamilton form is obviously the more fundamental
representation and the Lagrange form may even not exist. Furthermore
the
path integral derivation shows, while
the integral over the gauge field configuration
has to be taken with periodic boundary conditions
$A_i \lk x^0 = T , \vec{x} \rk =
A_i \lk x^0 = 0 , \vec{x} \rk $,
the integration over the electric fields is not constrained by
any temporal boundary condition.
}
\be
\label{G5}
Z [C'', C', J] & = & \int \cD \lk A_i, E_i \rk \pli_{\vec{x}}
\delta \lk f^a (A, E) \rk
\delta \lk \Gamma^a (A, E) \rk \pli_{{x}_0} Det \cM^{ab} (x_0)
\nonumber\\
& & \exp \left\{ \frac{i}{g^2} \int d^4 x \left[ E_i \partial_0 A_i -
\frac{1}{2} \lk E^{a}_{i}  E^{a}_{i} +  B^{a}_{i}  B^{a}_{i}  \rk - A_i
J_i \right] \right\} \hk . 
\ee
Here, $\cD \lk A_i, E_i \rk$ denotes the (flat) functional integral measure
over the gauge potential $A^{a}_{i}$ and the electric field $E^{a}_{i}$.
Furthermore, $f^a (A, E) = 0$ is 
the gauge fixing
constraint and 
$Det
\cM^{ab} (x,y)$, where $x_0 = y_0$, is the Faddeev-Popov
determinant. 
\bi

\no
In the following two sections we will explicitly resolve the Gau\ss \ law 
constraint $\delta \lk \Gamma^a \rk$ and the gauge constraint $\delta
\lk f^a \rk$ in (\ref{G5}) leaving a functional integral over unconstrained,
gauge fixed variables.
\bi

\section{Gauge fixing and partial resolution of Gau\ss' \hk law}
\bi

\no
Gau\ss ' law (\ref{30}) $\Gamma^a = 0$ has the generic form
\be
\vec{\nabla} \vec{E}^a = \rho^a \hk , \hs \rho^a = - \left[ A_i, E_i
\right]^a - J^{a}_{0} \hk ,
\ee
where $\rho^a$ is the total color charge density. Applying Gau\ss '
integration theorem it follows
\be
\oint\limits_{\partial M} d \vec{\Sigma} \vec{E}^a = Q^a \hs , \hs Q^a =
\int d^3 x \rho^a \hk .
\ee
For periodic electric fields the electric flux through the surface of
the box, $\oint d \vec{\Sigma} \vec{E}$, vanishes.
Consequently periodic boundary conditions to $E^{a}_{i} (x)$ can only
tolerate a vanishing total charge
\be
\label{46}
Q^a = 0 \hk . 
\ee
For the
resolution of Gau\ss ' law a proper choice of gauge fixing is
crucial. In the past a complete resolution of Gau\ss ' law has been
achieved in the gauge
$\epsilon^{a i k} E^{a}_{i} = 0 $
for $SU (2)$ in ref. \cite{goldstone}and an
extension to $SU (3)$ was considered in ref.  \cite{faddeev}. There have been
also attempts of a complete resolution of Gau\ss ' law in the Coulomb
gauge \cite{lee}. 
The Coulomb gauge, which is singled out in QED by the absence of
radiation of static charges has proved, however, to be inconvenient in
non-Abelian gauge theories, in particular for an explicit resolution of
Gau\ss ' law. In this respect axial type of gauges are much more
convenient as was already realized in refs. \cite{arnowitt}, \cite{schwinger}
and recently discussed in detail in ref. \cite{lenz1}, where an explicit
resolution of Gau\ss ' law in the canonical quantization approach has been
performed. Below we will perform an analogous resolution of Gau\ss ' law
in the functional integral approach. For this purpose it is convenient
to choose the 3-axis as the preferred direction of the axial gauge and
divide the Gau\ss ' law generator into parts parallel and perpenticular
to the 3-axis
\be
\label{48}
\Gamma (x) = \hat{D}_3 E_3 + \Gamma_\perp \hs , \hs  \Gamma_\perp =
\hat{D}_\perp E_\perp + J_0 \hk . 
\ee
If $ \hat{D}_3$ were regular the Gau\ss ' law $\Gamma = 0$ could be easily
resolved leading to an elimination of $E_3$. Unfortunately, as we will
explicitly see below, on the torus $\hat{D}_3$ has always zero modes,
independently of the used gauge. In fact, since $\hat{D}_3$ transforms gauge
covariantly its eigenvalues are independent of the gauge. Nevertheless
we can exploit the gauge freedom to cast $\hat{D}_3$ in
as simple a form as
possible. From this point of view the axial gauge $A_3 = 0$ would be
preferable. However,
this gauge condition conflicts with the
periodic boundary condition. This can be easily seen by considering the
Polyakov line operator
\be
\label{GX}
\cP_3 (x) = P \exp \lk \oint\limits_x d x'_3 A_3 \lk \bar{x}, x'_3 \rk \rk
\hk ,
\ee
where $P$ denotes path ordering and the integration runs from a 
point\footnote{Here and in the following we use the convention
$
(x) = \lk x^0, \vec{x} \rk = \lk \bar{x}, x_3 \rk \hk .$
}
$x
= \lk \bar{x}, x_3 \rk$ along the 3-axis to the point $x = \lk \bar{x},
x_3 + L \rk$. Due to the periodic boundary condition on $A_3$ the
integration in (\ref{GX}) runs over a closed loop but nevertheless due to
the path-ordering $\cP_3 (x)$ depends on the starting point $x$. Under
gauge transformation this quantity transforms as
\be
\cP_3 (x) \longrightarrow \cP^{\Omega}_{3} (x) = \Omega (x) \cP_3 (x)
\Omega^\dagger (x)
\ee
and one can obviously choose a gauge in which $\cP_3 (x)$ is diagonal
\be
\cP^{\Omega}_{3} (x) = e^{a_3 (x) L} \hs , \hs a_3 (x) = a^{c_0}_{3} T^{c_0}
\hk .
\ee
However, it is impossible to gauge transform $\cP_3 (x)$ to $\cP_3 (x) =
1$\footnote{
This can be also easily seen in the lattice
formulation.
Starting at $x_3 = 0$ one can bring the links $U_3 (x) = \exp \lk - a
A_3 (x) \rk$ to the gauge $U_3 (x) = 1$ except for the last link terminating
at $x_3 = L$, which cannot be gauged away  due to the periodic boundary
condition.}.
\bi

\no
For the resolution of Gau\ss ' law it is convenient to follow ref.
\cite{lenz1} and
use the
gauge
\be
\label{G16}
A^{ch}_{3} (x) = 0  \hs , \hs
{\rm or} \hs A^{\bar{a}}_{3} (x) = 0 \hk . 
\ee
This condition, of course, does not fix the gauge completely but allows
still for arbitrary abelian gauge transformations $\omega (x) \in H$. We
will later make use of this freedom. 
Let us also mention, that the gauge transformation necessary to bring a
given gauge field $A_i (x)$ into the form (\ref{G16}) requires in general also
topologically non-trivial gauge transformations.
\bi

\no
In the gauge (\ref{G16}) the operator $\hat{D}_3$ is block diagonal with
respect to the color neutral and charged components
\be
\label{54}
\hat{D}^{ab}_{3} = \lk
\begin{array}{cc}
\partial_3 \delta^{a_0 b_0} & 0\\
0 & 
\hat{D}^{\bar{a}
\bar{b}}_{3}
\end{array} \rk 
\ee
since $f^{a_0 b_0 c} = 0$.
Hence in this gauge the neutral part of the Gau\ss ' law
generator simplifies to
\be
\label{56a}
\Gamma^{a_0} (x) = \partial_3 E^{a_0}_{3} + \Gamma^{a_0}_{\perp} \hk .
\ee
On the space of periodic functions $ \xi_n (x)  =
e^{i
\omega_n x_3} , \omega_n = \frac{2 \pi n}{L}$
the operator $\partial_3$
has a zero eigenvalue $(n = 0)$ corresponding to a $x_3$-independent
eigenfunction. For simplicity of notation we have set here $L = L_3$.
The corresponding projection of $E^{n}_{3}$ onto this zero
mode
\be
\label{G34a}
e_3 (\bar{x}) = \frac{1}{L} \il^{L}_{0} d x_3 E^{n}_{3} \lk \bar{x}, x_3
\rk 
\ee
does not enter $\Gamma^n (x)$ (\ref{48}) and is hence not restricted by
Gau\ss \ law. 
\bi

\no
Since the eigenfunctions of $\partial_3$ belonging to zero and non-zero 
eigenvalues are orthogonal in
the Hilbert space of periodic functions, the neutral part of the Gau\ss '
law constraint separates in the two independent constraints
corresponding to the subspaces
of the zero and non-zero eigenvalues. Defining 
\be
\label{G23}
\gamma_\perp = \frac{1}{L} \il^{L}_{0} d x_3 \Gamma^{n}_{\perp} \hs , \hs 
\Gamma'^{n}_{\perp} = \Gamma^{n}_{\perp} - \gamma_\perp  \hk , 
\ee
we have
\be
\label{G22}
\delta (\Gamma^n) = \delta \lk \partial_3 E'^{n}_{3} + \Gamma'^{n}_{\perp} \rk
\delta \lk \gamma_{\perp} \rk = \delta \lk \partial_3 E'^{n}_{3} +
\Gamma^{n}_{\perp} \rk \delta \lk \gamma_\perp \rk \hk , 
\ee
where 
\be
E'_{3} = E_3 - e_3 
\ee
lives entirely in the subspace of eigenfunctions
with non-zero eigenvalues of $\partial_3$. The constraint of the first
$\delta$-function can be easily resolved. Defining by $ \partial'_3$ the
operator resulting from $\partial_3$ when the zero eigenvalue is 
removed, 
we obtain (with
$\partial_3 E'^{n}_{3} = \partial'_3 E'^{n}_{3} $)
\be
\label{62a}
\delta \lk \partial'_3 E'^{n}_{3} + \Gamma^{n}_{\perp} \rk = \frac{1}{\det
\partial'_3} \delta \lk E'^{n}_{3} + \frac{1}{\partial'_3} \Gamma^{n}_{\perp}
 \rk
\hk . 
\ee
\bi

\no
Hence the neutral part of Gau\ss \ law eliminates the variable
$E'^{n}_{3}$. In addition we now exploit the residual invariance
under Abelian gauge 
transformations to remove also the corresponding conjugate field
variable
\be
\label{G34b}
A'_3 = A_3 - a_3 \hs , \hs a_3 = \frac{1}{L} \int d x_3 A^{n}_{3} \hs
\ee
by imposing the gauge condition
\be
\label{G24}
\partial_3  A^{n}_{3} (x) = 0 \hk . 
\ee
Since $\partial_3 A'^{n}_{3} = \partial'_3  A'^{n}_{3}$ this gauge
implies $ A'^{n}_{3} (x) = 0$ and hence leaves from $ A^{n}_{3} (x)$
only the $x_3$-independent part $a_3 \lk \bar{x} \rk$.
\bi

\no
By construction (see eqs. (\ref{G34b}) and (\ref{G34a})) the reduced
Abelian fields $a_3 \lk \bar{x} \rk$ and $e_3  \lk \bar{x} \rk$ are
canonically conjugated variables.
Note also, that the change of variables from $A^{n}_{3}$ to $\lk
A'^{n}_{3}, a_3 \rk$ (and correspondingly from $E^{n}_{3}$ to $\lk
E'^{n}_{3}, e_3 \rk$) does not yield any non-trivial Jacobian since
$A'^{n}_{3}$ and $a_3$ are orthogonal coordinates in the sense that they
belong to orthogonal subspaces of the Hilbert space of periodic eigenfunctions 
of
$i \partial_3$.
\bi

\no
The gauge condition (\ref{G24}) has also the advantage that it
enormously simplifies the operator (\ref{54})
\be
\hat{D}^{\bar{a} \bar{b}}_{3} = \delta^{\bar{a} \bar{b}} \partial_3 +
\hat{a}^{\bar{a} \bar{b}}_{3} \lk \bar{x} \rk =: \hat{d}^{\bar{a}
\bar{b}}_{3}
\ee
which enters the charged part of Gau\ss \ law (\ref{48})
\be
\Gamma^{\bar{a}} =  \hat{d}^{\bar{a}
\bar{b}}_{3}
 E^{\bar{b}}_{3} + \Gamma^{\bar{a}}_{\perp} \hk .
\ee
(The evaluation of the eigenvalues, and hence the inversion of
$\hat{d}_3$ becomes trivial since $a_3 \lk \bar{x} \rk$ is independent
of $x_3$, see below.) Let us also mention that equation (\ref{G16}) and
(\ref{G24}) define a variant of 't Hooft's maximal Abelian gauge
\cite{thooft}, which preserves invariance under $x_3$-independent
Abelian gauge transformation.
\bi

\no
For the time being, let us assume that $\hat{d}^{a \bar{b}}_{3}$ 
has 
no zero eigenvalue in the charged subspace. (We will later see that the
system dynamically avoids configurations $a_3 (\bar{x}) = 0$ giving
rise to zero modes of $\hat{d}_3$.)
The charged part of the Gau\ss ' law can now be used to eliminate the
charged part of $E_3$ by using 
\be
\label{G31}
\delta \lk \Gamma^{\bar{a}} \rk = \frac{1}{\det \hat{d}_{3}}
\delta \lk E^{\bar{a}}_{3} + \lk \hat{d}^{-1}_{3} \rk^{\bar{a}
\bar{b}} \Gamma^{\bar{b}}_{\perp} \rk  \hk .
\ee
Later we will observe that the corresponding functional determinant 
$\det \hat{d}^{\bar{a}
\bar{b}}_{3}$ will be cancelled by the Faddeev-Popov
determinant.
\bi

\no
We can use now the two constraints (\ref{62a}) and (\ref{G31}) arising 
from the Gau\ss '
law to
integrate out explicitly the electric field variables $E'^{c_0}_{3} ,
E^{\bar{a}}_{3}$ leaving from $E_3$ only the $x_3$-independent
neutral part $e_3$ (\ref {G34a}). 
Furthermore the two gauge constraints (\ref{G16}) and (\ref{G24}) 
eliminate the gauge
variables $A^{ch}_{3}$ and $A'^{n}_{3}$ respectively, leaving from the
gauge potential $A_3$ only the neutral $x_3$-independent part $a_3
(\bar{x})$. 
Since the changes of
variables from $E^{c_0}_{3}$ to $E'^{c_0}_{3} , e_3$ and analogously
from $A^{c_0}_{3}$ to $A'^{c_0}_{3}, a_3$ are trivial, i.e.
the corresponding jacobians equal one. We then obtain  from (\ref{G5})
\be
Z [J] & = & \int \cD \lk A_\perp, a_3, E_\perp, e_3 \rk \pli_x
\delta \lk \bar{f}^{c_0} (A) \rk \delta \lk \gamma^{n}_{\perp} \rk \pli_{x^0}
Det \cM^{ab} \cdot \left[ Det \lk \partial'_3 \rk
Det \hat{d}^{\bar{a} \bar{b}}_{3} \right]^{-1}\nonumber\\
& & \exp \left\{ \frac{i}{g^2} \left[ L \int d^3 \bar{x} \lk e_3
\partial_0 a_3 - \frac{1}{2} e_3 \lk \bar{x} \rk e_3  \lk \bar{x} \rk
\rk \right] + \int d^4 x E^{a}_{\perp} \partial_0 A^{a}_{\perp}\right.
 \nonumber\\
& & \left. - \frac{1}{2}
\int d^4 x \left[ \lk \hat{d}^{-1}_{3} (ch) \Gamma^{ch}_{\perp} \rk^2 +
\lk \partial'^{- 1}_{3} \Gamma'^{n}_{\perp} \rk^2 + E_\perp 
E_\perp + B^2 \right] \right\}
\hk .
\ee
\bi

\no
Here $\delta \lk \bar{f}^{c_0} (A) \rk$ denotes the gauge condition
necessary to fix the residual invariance under $x_3$-independent Abelian
gauge transformations, which is left by the constraints (\ref{G16}) and
(\ref{G24}). This residual gauge will be fixed in the following section
when we resolve the residual Gau\ss \ law $\gamma_\perp = 0$.
\bi

\section{Resolution of the residual Gau\ss ' law}
\bi

\no
The residual Gau\ss \ law constraint
\be
\label{G36}
\gamma_\perp = \frac{1}{L} \int d x_3 \lk \nabla_\perp E^{n}_{\perp} +
\left[ A_\perp, E_\perp \right]^n + J^{n}_{0} \rk 
\ee
can be used to remove the $x_3$-independent part of $ E^{n}_{\perp}$
which is longitudinal in the 1-2-plane (\underline{i} = 1,2) defined by
\be
\label{58}
e_\perp := \nabla_\perp \frac{1}{\Delta'_\perp} \frac{1}{L} \int d x_3
\nabla_\perp E^{n}_{\perp} \equiv l E^{n}_{\perp} \hk ,
\ee
where $\triangle'_\perp$ is  the two
dimensional laplacian, $\nabla_\perp \nabla_\perp$, in the Hilbert space
of periodic functions with the zero mode
omitted. Its inverse is defined in the space of periodic functions by
the Green's function
\be
\label{56}
G^{(2)} \lk x''_\perp, x'_\perp \rk = \langle x''_\perp |
\frac{1}{- \triangle'_\perp} | x' \rangle =
\frac{1}{(2 \pi)^2} \sli_{\vec{n}_\perp \neq 0 }
\frac{1}{\vec{n}^{2}_{\perp}}
e^{i \vec{n}_\perp \cdot \lk \vec{x}''_\perp - x'_\perp \rk \frac{2
\pi}{L}} \hk , \hk
n_\perp =
\lk n_1, n_2 \rk 
\ee 
which obviously satisfies periodic boundary conditions.
Note that the longitudinal projector $l$ defined by eq. 
(\ref{58}) is in fact an
orthogonal projector, $l \cdot l = l$. This follows from the relation
\be
\label{G59}
\int d^2 x^{''}_{\perp} \langle x | \triangle_\perp | x'' \rangle
\langle
x'' | \frac{1}{\triangle'_\perp} | x' \rangle = \delta^{(2)} \lk x_\perp
- x'_\perp \rk - \frac{1}{L^2} \hk , 
\ee
where 
\be
\delta^{(2)} \lk x_\perp, y_\perp \rk = \frac{1}{L^2}
\sli_{\vec{k}_\perp} e^{{i \vec{k}_\perp \lk
\vec{x}_\perp - \vec{y}_\perp \rk}} \hs , \hs \vec{k}_\perp = \lk
\frac{2 \pi}{L_1} n_1, \frac{2 \pi}{L_2} n_2 \rk 
\ee
is the 2-dimensional periodic $\delta$-function $\lk \delta^{(2)} \lk
x_\perp + e_{i} L_i, y_\perp \rk  = \delta^{(2)} \lk x_\perp, y_\perp \rk
\rk$ and the last term in (\ref{G59}) arises from the fact that in
$\triangle'_\perp $ (\ref{56})  
the zero mode $n_1 = n_2 = 0$ is excluded. This term,
however, does not
contribute when $l$ acts on vector fields $V_i (x)$
 periodic in $x_1$ and $x_2$. In fact, from the definition of the
longitudinal field $e_\perp$ (\ref{58}) we find by using (\ref{G59})
\be
\label{C4a}
\nabla_\perp e_\perp = \frac{1}{L} \int d x_3 \nabla_\perp E^{n}_{\perp}
- \frac{1}{L {L_1 L_2}} \int d^3 x \nabla_\perp E^{n}_{\perp} \hk , 
\ee
where the last term vanishes for periodic electric fields, so
that we obtain
\be
\label{G61}
\nabla_\perp e_\perp = \frac{1}{L} \int d x_3 \nabla_\perp E^{n}_{\perp}
\hk . 
\ee
The residual Gau\ss ' law (\ref{G36}) then simplifies to
\be
\label{X75}
\gamma_\perp = \nabla_\perp e_\perp - \rho^{(2)} = 0 \hk , 
\ee
where
\be
\label{62}
 \rho^{(2)} = - \frac{1}{L} \int d x_3 \lk \left[ A_\perp, E_\perp
\right]^n + J^{n}_{0} \rk \hk . 
\ee
Since by definition, $e_\perp$ (\ref{58}) is a curl-free, 2-dim. 
vector field,
$\nabla_\perp \times e_\perp = 0$, it has the representation
\be
\label{C1a}
e_\perp = - \nabla_\perp \varphi \lk x_\perp \rk \hk , 
\ee
where the scalar potential $\varphi (x)$ follows from the residual
Gau\ss ' law (\ref{X75}) 
\be
\label{C2a}
\varphi \lk x_\perp \rk = \int d^2 y_\perp G^{(2)} \lk x_\perp, y_\perp
\rk \rho^{(2)} \lk y_\perp \rk \hk . 
\ee
In fact, inserting (\ref{C2a}) into (\ref{C1a}) and taking the divergence
we find with the help of (\ref{G59})
\be
\nabla_\perp e_\perp = \rho^{(2)} - \bar{\rho} \hs , \hs \bar{\rho} =
\frac{1}{L_1 L_2} \int d^2 x_\perp  \rho^{(2)} = \frac{1}{L {L_1 L_2}} 
Q^n \hk ,
\ee
where $Q^n$ is the total charge (in the Cartan subgroup), which 
according to (\ref{46}) has to vanish
for periodic $E^{a}_{i}$-fields, so that $\bar{\rho} = 0$ and (\ref{C1a})
solves, in fact, (\ref{X75}). 
\bi

\no
For non-vanishing total charge $Q^n \neq 0$
Gau\ss ' law requires one to abandon the periodic boundary
condition to the 
electric fields and the second term in (\ref{C4a}) no longer vanishes.
Even in this case eq. (\ref{X75}) is still solved by eqs.  (\ref{C1a}),
(\ref{C2a}).
Therefore the resolution of the residual part of Gau\ss ' law leads to
the elimination of the longitudinal part $e_\perp$ of the neutral vector field
${E}^{n}_{\perp}$ and we are left with the transversal part
\be
\label{39}
E'_\perp = E_\perp - e_\perp 
\ee
as dynamical quantity.
\bi

\no
Note, that only the charged parts $A^{ch}_{\perp}$ and $E^{ch}_{\perp}$ enter
$\rho^{(2)}$ (\ref{62}) and thus $e_\perp$. Furthermore, $e_\perp$ and
$E'^{n}_{\perp}$ live in orthogonal subspaces of the Hilbert space of
periodic functions in $x_3$. Therefore the change of 
variables from $\lk E^{n}_{\perp},  E^{ch}_{\perp} \rk$ to $\lk
{E}^{n}_{\perp} =  E'^{n}_{\perp} + e_\perp,  E^{ch}_{\perp} \rk$ does
not give rise to any non-trivial Jacobian. Let us also emphasize that
after resolution of Gau\ss \ law (\ref{X75}) 
$e_\perp$ is not an integration variable but a function of $ E^{ch}_{\perp},
A^{ch}_{\perp}$ and independent of the remaining integration variables $
{E'}^{n}_{\perp},  {A}^{n}_{\perp}$ etc.
\bi

\no
We can exploit now the residual invariance under $x_3$-independent
Abelian gauge transformations, left by the constraints (\ref{G16}) and
(\ref{G24}), to remove the field
\be
\label{E65}
a_\perp \lk \bar{x} \rk = \lk l A^{n}_{\perp} \rk \lk \bar{x} \rk
\ee
canonically conjugated to $e_\perp$. Since by definition of the
longitudinal projector $l$ (\ref{58}) this field is curl-free,
$\vec{\nabla}_\perp \times a_\perp = 0$ and, for periodic fields $A_\perp
(x)$, satisfies the relation (c.f. eq. (\ref{G61}) )
\be
\nabla_\perp a_\perp \lk \bar{x} \rk = \frac{1}{L} \int d x_3
\nabla_\perp A^{n}_{\perp}
\ee
it suffices to require the gauge
\be
\label{G26}
\frac{1}{L} \int d x_3 \nabla_\perp A^{n}_{\perp} \lk \bar{x}, x_3 \rk =
0
\ee
to make $a_\perp$ vanishing
\be
a_\perp = 0 \hk .
\ee
In the following we will denote by $A'_\perp$ the field satisfying the
gauge condition (\ref{G26}), i.e.
\be
\label{G39}
A'_\perp = A_\perp - a_\perp \hk . 
\ee
Since $e_\perp$ and $a_\perp$ live in the Cartan subgroup we can
trivially extend eqs. (\ref{39}) and (\ref{G39}) to the charged field
components, where they read
\be
E'^{ch}_{\perp} = E^{ch}_{\perp} \hs , \hs A'^{ch}_{\perp} =
A^{ch}_{\perp} \hk .
\ee
We can then express $e_\perp$ defined by (\ref{C1a}) and (\ref{C2a}) as
\be
\label{Z1}
e_\perp \lk \bar{x} \rk = \nabla^{x}_{\perp} \frac{1}{L} \int d^3 y
G^{(2)} \lk x_\perp, y_\perp \rk \lk \left[ A'_\perp, E'_\perp \right]^n
+ J^{n}_{0} \rk \lk x^0, \vec{y} \rk \hk . 
\ee
(Note that only the charge fields $ E^{ch}_{\perp}, A^{ch}_{\perp}$
enter the commutator).
\bi

\no
Similarly we can express the Gau\ss \ law generator $\Gamma_\perp$ (\ref{48}) in
terms of the new variables
\be
\label{E72}
\Gamma^{n}_{\perp} & = & \nabla_\perp \lk E'^{n}_{\perp} + e_\perp \rk +
\left[ A'_\perp, E'_\perp \right]^n + J^{n}_{0} \hk , \nonumber\\
\Gamma^{ch}_{\perp} & = & \nabla_\perp E'^{ch}_{\perp} + \left[
A'_\perp, E'_\perp + e_\perp \right] + J^{ch}_{0} \hk . 
\ee
Furthermore since $e_\perp \lk \bar{x} \rk$ (\ref{Z1}) is independent of
$x_3$, i.e. $\partial_3 e_\perp = \partial'_3 e_3 = 0$, it drops out
from
\be
\partial'^{-1}_{3} \Gamma^{n}_{\perp} \equiv \frac{1}{\partial'^{2}_{3}}
\partial'_3 \Gamma^{n}_{\perp} \hk .
\ee
and the neutral part of the Gau\ss \ law generator (\ref{E72}) can be
replaced by
\be
{\Gamma}^{c_0}_{\perp} & = & \nabla_\perp E'^{c_0}_{\perp} + 
\left[ A'_\perp,
E'_\perp \right]^{c_0} + J^{c_0}_{0} 
\ee
Recall, that the change of integration variables from ${E}^{n}_{\perp}
\to \lk E'^{n}_{\perp} , e_\perp \rk$ yields a trivial Jacobian equal to
one since $E'^{n}_{\perp}$ and $e_\perp$ are orthogonal components of
${E}^{n}_{\perp}$ in the Hilbert space of periodic functions in
$x_3$. The same is true for the change of variables from $A^{n}_{\perp}$
to $\lk A'^{n}_{\perp}, a_\perp \rk$. 
Therefore, after complete resolution of Gau\ss ' law and
implementation of the gauge fixing contraints, we are left with the
following functional integral representation of Yang-Mills theory
\be
\label{G80}
Z & = & \int \cD \lk E'_\perp , e_3, A'_\perp , a_3 \rk
\pli_{x_0} Det \cM^{ab} Det^{-1} \lk 
 \hat{d'}_{3} \rk \nonumber\\
& & \exp \left\{ \frac{i}{g^2} \left[ L \int d^3 \bar{x} \lk e_{3}
\partial_0 a_{3}  - \frac{1}{2}
e_{3}
e_{3}
 \rk + \int d^4 x E'_{\perp} \partial_0
A'_{\perp} + \int \lk a_3 J_3 + A'_\perp J'_\perp \rk 
\right. \right. \nonumber\\ 
& & \left. - \frac{1}{2} \int  \left[ \lk  \hat{d'}^{-1}_{3}
 \Gamma_{\perp} \rk^2 + 
 E'_{\perp}  E'_{\perp} + 
e^{2}_{\perp} + 
\lk B(A') \rk^2 \right] \right\}
\hk , 
\ee
where the residual abelian gauge constraint (\ref{G26}) has been used to
replace the perpendicular field $A_\perp$ by its two-dimensional-transversal
part $A'_\perp$, see eq. (\ref{G39}). Furthermore, the
magnetic field $B(A')$ is defined in terms of the reduced field variables
due to the implementation of Gau\ss ' law by
\be
F_{\underline{i} 3} & = & \left[ D'_{\underline{i}} , d_3 \right] \hs ,
 \hs d_3 = \partial_3 + a_3 \hk , \hk d'_3 = \partial'_3+a_3
\nonumber\\
F_{\underline{i} \underline{j}} & = &  \left[ D'_{\underline{i}} , D'_j
\right] \hs , \hs D'_i = \partial_i + A'_i 
\ee
Let us also emphasize, that there are
no cross terms between the reduced electric field $ E'_\perp$ and the
static electric field $e_\perp$. This is a consequence of $\int d x_3
\nabla_\perp E'_\perp = 0$, which holds due to 
the periodicity of the fields.
\bi

\no
It remains to calculate the Faddeev-Popov determinant which is done in
the next section.
\bi

\newpage

\section{Evaluation of the Faddeev-Popov determinant}
\bi

\no
For the above chosen gauge the Faddeev-Popov determinant is straight
forwardly evaluated.
The two abelian gauge fixing conditions (\ref{G24}) and (\ref{G26}) are
independent of each other, i.e. they belong to orthogonal subspaces of
the Hilbert space of periodic functions in $x_3 \in \left[ 0, L
\right]$. Both conditions can therefore be absorbed into a single gauge
constraint for the neutral component of the gauge field
\be
\label{G71}
f^{a_0} (x) = \partial_3 A^{a_0}_{3} \lk \bar{x} , x_3 \rk +
\nabla_\perp
\frac{1}{L} \il^{L}_{0} d x_3 A^{a_0}_{\perp} \lk \bar{x} , x_3 \rk \hk
.
\ee
\bi

\noindent
Furthermore the gauge (\ref{G16}) defines a color charged gauge
functional
\be
\label{GG2}
f^{\bar{a}} = A^{\bar{a}}_{3} \hk . 
\ee
For the above gauge functionals (\ref{G71}) and (\ref{GG2}) 
the Faddeev-Popov kernel
$
\cM^{ab} (x, y) 
$
becomes $(x_0=y_0)$
\be
\cM^{a b_0} (x, y) & = & \hat{D}^{a b_0}_{3} (x) \nabla^{y}_{3}
\delta^{(3)}
\lk \vec{x} - \vec{y} \rk + \frac{1}{L}  \hat{D}^{a b_0}_{\perp} (x)
\nabla^{y}_{\perp}  \delta^{(2)} \lk x_\perp - y_\perp \rk  \nonumber\\
\cM^{a \bar{b}} (x, y) & = & \hat{D}^{a \bar{b}}_{3} (x) \delta^{(3)}
\lk \vec{x} - \vec{y} \rk \hk .
\ee
This expressions hold so far for arbitrary gauge field configurations.
We
need, however, these expressions only on the gauge manifold, i.e. for
those field configurations which fulfill the above chosen gauge
constraints. Using $f^{a b_0 c_0} = 0$, which implies $\hat{a}^{a
b_0}_{3} = 0$ the Faddeev-Popov kernel reduces at the chosen gauge
orbits to
\be
& &
\cM^{ab} (x, y)
  \equiv 
\lk  \begin{array}{cc}
\cM^{a_0 b_0} & \cM^{a_0 \bar{b}} \\
\cM^{\bar{a} b_0} & \cM^{\bar{a} \bar{b}}
\end{array} \rk \\
& = &
\lk \begin{array}{cc}
- \delta^{a_0 b_0} \lk \nabla^{x}_{3}  \nabla^{x}_{3} \delta^{(3)}
\lk \vec{x} - \vec{y} \rk + \frac{1}{L} \nabla^{x}_{\perp}
\nabla^{x}_{\perp} \delta^{(2)} \lk x_\perp - y_\perp \rk \rk
& 0 \\
- \frac{1}{L} \hat{A}^{\bar{a} b_0}_{\perp} (x) \nabla^{x}_{\perp}
\delta^{(2)} \lk x_\perp - y_\perp \rk & 
\hat{d}^{\bar{a} \bar{b}}_{3} (x) \delta^{(3)} \lk \vec{x} - \vec{y} \rk
\end{array} \rk \hk . \nonumber
\ee
Since this matrix has triangle form, we find for the Faddeev-Popov
determinant finally
\be
\label{D1a}
Det \cM^{ab} (x, y) & = &  Det \left[ - \delta^{a_0 b_0} \lk
\nabla^{x}_{3} \nabla^{x}_{3} \delta^{(3)} \lk \vec{x} - \vec{y} \rk +
\frac{1}{L}  \nabla^{x}_{\perp}  \nabla^{x}_{\perp} \delta^{(2)} \lk
x_\perp - y_\perp \rk \rk \right]
\nonumber\\
& & Det \lk \hat{d}^{\bar{a} \bar{b}}_{3}
\delta^{(3)}
\lk \vec{x} - \vec{y} \rk \rk \hk . 
\ee
It factorizes into contributions arising from the Cartan subgroup
(first factor) and the coset space.
The former one is an
irrevelant constant and will be dropped in the following. The
contribution from the coset space can be easily calculated since the
eigenvalues of $ \hat{d}^{\bar{a} \bar{b}}_{3}$ are analytically known,
see Appendix A. But
for the moment we do not need the explicit form of $Det \hk \hat{d}_3$.
\bi

\no
A glance at eq. (\ref{D1a}) shows that (the non-trivial part of) the
Faddeev-Popov determinant cancels precisely the determinant $\lk Det
\hat{d}_3 \rk^{- 1}$ arising from the resolution of Gau\ss ' law.
Consequently eq. (\ref{G80}) reduces to
\be
\label{G81}
Z & = & \int \cD \lk E'_\perp, e_3, A'_\perp, a_3 \rk \exp \left\{
\frac{i}{g^2} \left[ L \int d^3 \bar{x} \lk e_3 \partial_0 a_3 -
\frac{1}{2} e_3 e_3 \rk \right. \right.   \nonumber\\
& & \left. \left. + \int d^4 x E'_\perp \partial_0 A'_\perp -
\frac{1}{2} \int d^4 x \left[ \lk \hat{d}'^{-1}_{3} \Gamma_\perp \rk^2 +
E'_\perp E'_\perp + e^{2}_{\perp} + \lk B' (A) \rk^2 \right] \right] \right\}
. 
\ee
This is the desired functional integral representation of Yang-Mills
theory in uncon\-strained, gauge-fixed variables, resulting from a
complete resolution of Gau\ss \ law.  Note that in the unconstrained 
theory the functional integration
over the canonical variables is performed with flat integration measure.
(There is no preexponential factor, e.g. a functional determinant,
which could be interpreted as non-trivial measure.) This is obviously a
general feature of Yang-Mills theory in unconstrained variables (provided
one chooses a gauge condition which is canonically conjugated to the
Gau\ss \ law  constraint) and could, perhaps, have been anticipated in view of
the fact that the Faddeev-Popov kernel is given by $ \cM^{ab} (x,y) = 
\left\{ f^a
(x), \Gamma^b (y) \right\}$, where $\{ , \}$ denotes the Poisson
bracket.
\bi

\no
The cancelation of the Faddeev-Popov determinant against the determinant
arising from the resolution of Gau\ss ' law was also obtained in ref.
\cite{goldstone}, where the gauge was fixed by demanding that the
antisymmetric part of the matrix $E^{a}_{i}$ vanishes.
In that case Gau\ss ' law requires the vanishing of the antisymmetric part
of $A^{a}_{i}$ and one ends up with a functional integral over the symmetric
parts of $A^{a}_{i}$ and $E^{a}_{i}$, where unfortunately the remaining
electric field variables cannot explicitly been integrated out. In this
respect the present approach has the advantage over refs.
\cite{goldstone,faddeev} in that the remaining unconstrained electric
field variables in (\ref{G81}) can still be integrated out in closed
form. This will be done in section 7.
\bi

\no
Before concluding this section let us notice that {\it assuming} a 
flat integration
measure the functional integral representation (\ref{G81}) could have
also been derived 
by starting from the Yang-Mills Hamilton operator in
unconstrained variables obtained in ref. \cite{lenz1} in the canonical
operator approach and following the standard procedure \cite{hibbs}.
In this sense the present functional integral derivation of the
unconstrained Yang-Mills theory (\ref{G81}) is equivalent to
the canonical operator approach of ref. \cite{lenz1}. We believe, however,
that the functional integral representation derived in the present paper
(\ref{G81}) is more flexible than the operator approach when it comes to
an approximate solution of the theory.
\bi

\no
Finally a comment on the gauge-fixing is in order. We have fixed the
gauge in such a way to remove the components of the gauge field $A_i
(x)$ which are canonically conjugate to those components of the
electric field $E_i (x)$ which are eliminated by Gau\ss \ law.  This has
lead to the gauge conditions (\ref{G16}), (\ref{G26}) and (\ref{G24}),
which eliminate $A^{ch}_{3}, a_\perp$ and make $A^{n}_{3} = a^{n}_{3} $
independent of $x_3$. These gauge constraints do, however, not yet fix the
gauge completely but leave a residual gauge invariance which consist of
i) (global) permutations of the color indices of the fundamental
representations, i.e. elements of the Weyl (sub-) group $S_N$ of the
gauge group $SU (N)$, ii) global Abelian gauge transformations and iii)
displacement transformations $\Omega = e^{- \vec{\alpha} \vec{x}}$ with
$\vec{\alpha}$ an arbitrary but fixed c-number 3-vector. These residual
gauge symmetries were also found in ref. \cite{lenz1}. For completeness we
work out the emergence of these residual gauge symmetries in the present
functional integral approach in Appendix C.
\bi


\section{The Wilson loop}
\bi

\noindent
Below we evaluate the potential between two static color charges or,
equivalently, the Wilson loop
\be
\label{7.1}
W (C) = \langle Tr P \exp \lk - \oint d x_\mu A_\mu (x) \rk \rangle \hs
\ee
in the (Euclidean version of) the gauge-fixed theory defined by (\ref{G81}) . 
For simplicity we consider a planar rectangular Wilson loop $C$, which
by Lorentz invariance can be placed into the $0 - 3$ plane. One should
note here, however, that the present approach (\ref{G81}) has not been
formulated in a manifestly Lorentz covariant way, although all Green
functions (calculated in the full theory) will respect Lorentz
invariance. As a consequence the quality of approximations will depend
in general on the chosen Lorentz frame. 
\bi

\no
The present approach obviously singles out the $0-$ and $3-$ axis. (It
starts from the Weyl gauge and eliminates most of the degrees of freedom
of $A_3$ and $E_3$ by gauge-fixing and resolution of Gau\ss ' law,
respectively.) We therefore expect that the Wilson loop is most
efficiently evaluated when placed in the $0-3$-plane. Then the
$A'_\perp$ field will not explicitly enter the Wilson loop. Therefore we
will ignore it together with its conjugate variable $E'_\perp$ 
since we anyhow expect the dominant infrared behaviour to
be governed by the $a_3 (\bar{x}), e_3 (\bar{x})$ fields. 
The generating functional of
axial-gauge fixed Yang-Mills theory (\ref{G81}) reduces then to
\be
\label{S1}
Z [J] & = & = \int \cD \lk a_3, e_3 \rk \exp \left\{ \frac{i}{g^2}
\left[ L \int d^3 \bar{x} \lk e_3 \partial_0 a_3 - \frac{1}{2} e_3 e_3
\rk + \int a_3 \right.  \right. \nonumber\\
& & \left. \left. - \frac{1}{2} \int d^4 x \left[ J_0 
\frac{1}{- \hat{d}'_3  \hat{d}'_3}
J_0 + \lk e^{0}_{\perp} \rk^2 \right] \right] \right\} \hk , 
\ee
where
\be
\label{106}
e^{(0)}_{\perp} = e_\perp |_{E_\perp = 0} = - \nabla_\perp \frac{1}{L}
\int d^3 y G^{(2)} \lk x_\perp, y_\perp \rk J^{n}_{0} (y) \hk . 
\ee
The last two terms describe the interaction between static charges
$J_0$. The last term can be cast into the form
\be
\label{S2}
\int \lk e^{(0)}_{\perp} \rk^2 = - \int d x_0 d^2 x_{\perp}  d^2 y_\perp
\bar{J}^{c_0}_{0}  \lk x_0, x_\perp \rk G^{(2)} \lk x_\perp, y_\perp \rk
\bar{J}^{c_0}_{0}  \lk x_0, y_\perp \rk \hk , 
\ee
where a partial integration has been performed and
\be
\label{S3}
\bar{J} = \frac{1}{L} \int d x_3 J_0 (x) \hk .
\ee
This quantity obviously vanishes in the thermodynamic (infinite volume)
limit $L \to \infty$ for any localized charge distribution $J_0 (x)$.
To illustrate the meaning of this term let us consider two opposite
Abelian charges $\lk q, - q \rk$ separated by a distance $R$. If we
place these two charges on a line parallel to the 3-axis, e.g.
\be
\label{S4}
{J}^{c_0}_{0}  \lk x \rk = q^{c_0} \delta \lk x_1 \rk \delta \lk x_2 \rk
\left[ \delta \lk x_3 - \frac{R}{2} \rk - \delta \lk x_3 + \frac{R}{2}
\rk \right] 
\ee
then obviously $\bar{J}_{0}  \lk \bar{x} \rk = 0$ and this term does not
contribute. But it does contribute when we place the charges in the
$x-y$ plane, e.g.
\be
\label{S5}
{J}^{c_0}_{0}  \lk x \rk = q^{c_0} \delta \lk x_3 \rk \delta \lk x_2 \rk
\left[ \delta \lk x_1 - \frac{R}{2} \rk - \delta \lk x_1 + \frac{R}{2}
\rk \right] \hk .
\ee
If we take, for simplicity, the thermodynamic limit $L_1, L_2 \to
\infty$ of $G^{(2)} \lk x_\perp,
y_\perp \rk$ (\ref{56})
\be
\label{S6}
G^{(2)} \lk {x}_\perp , {y}_\perp \rk = \ln | \vec{x}_\perp -
\vec{y}_\perp | 
\ee
we receive from eq. (\ref{S2}), besides an infinite constant, 
a logarithmically increasing potential.
However, we do not expect that eq. (\ref{S1}), which discards all
perpendicular degrees of freedom $A'_\perp, E'_\perp$, can give a
realistic description of the interaction between two charges sitting in
the 1-2-plane. As discussed before the present approach singles out the
3-axis and in fact preserves the rotational symmetry around the 3-axis.
Let us therefore consider the axial symmetric charge distribution
(\ref{S4}). In this case $e^{(0)}_{\perp} = 0$, and from the
second to last term in (\ref{S1})
we obtain the static interaction potential
\be
\label{S7}
V = \frac{1}{2 g^2} \lk \delta (0) \rk^2 q^2 G^{(1)} (R, 0) \hk , 
\ee
where $ G^{(1)}$ is the Green's function of $- \partial'^{2}_{3}$. If we
again take the thermodynamic limit $L \to \infty$
\be
\label{S8}
 G^{(1)} \lk x_3, y_3 \rk = \langle x_3 | \frac{1}{- \partial^{2}_{3}} |
y_3 \rangle = \frac{1}{2} |x_3 - y_3 | 
\ee
we obtain a linearly raising potential
\be
\label{S9}
V = \sigma R 
\ee
with a string tension
\be
\label{S10}
\sigma = \frac{q^2}{g^2} \lk \delta (0) \rk^2 \hk . 
\ee
Here it is understood that $\delta (0)$ is regularized in an
appropriate way. 
The above obtained interaction potential is in agreement
with the findings of the canonical quantization approach
\cite{lenz1}, see also ref. \cite{moniz}.
\bi

\no
The emergence of the linear confinement potential  in the 3-direction
should come as no surprise since, except for  the dummy $x_1,
x_2$-dependence, eq. (\ref{S1}) represents the generating functional for
$1 + 1$ dimensional Yang-Mills theory, which is known to confine. In
fact, if we ignore the $e^{(0)}_\perp$ term (which, as seen above,
vanishes in the thermodynamic limit $L_3 \to \infty$ for any localized
charge distribution $J_0 (x)$ and furthermore depends only on the
``dummy'' coordinates $x_1,
x_2$) and linearize the term quadratic in $J_0$ by means of a field $a_0
(x)$, and furthermore perform the integration  over $e_3$ we obtain
\be
\label{S11}
Z [J] & = & \int \cD \lk a_0, a_3 \rk \exp \left\{ \frac{i}{g^2} \int
d^2 x_\perp \left[ \int d x_0 d x_3 \lk \frac{1}{2} \lk \partial_0 a_3
\rk^2 \right. \right. \right. \nonumber\\
& & \left. \left. \left. + \frac{1}{2} \lk \hat{d}_3 a_0 \rk^2 + 
a_0 J_0 + a_3 J_3 \rk
\right] \right\} \hk . 
\ee
Here the first two terms in the bracket combine to $f^{2}_{\bar{\mu}
\bar{\nu}}$, where
\be
f_{\bar{\mu}
\bar{\nu}} = \partial_{\bar{\mu}} a_{\bar{\nu}} - \partial_{\bar{\nu}}
a_{\bar{\mu}} \hs , \hs \bar{\mu}, \bar{\nu} = 0,3
\ee
and additionally $a_3$ satisfies, by construction (see.
eqs. (\ref{G16}), (\ref{G34b})), the gauge
\be
a^{ch}_{3} = 0 \hs , \hs \partial_3 a^{n}_{3} = 0 \hk .
\ee
In $D = 1 + 1$ the corresponding Faddeev-Popov determinant is an
irrelevant constant. Thus eq. (\ref{S11}) represents in fact the
properly gauge-fixed generating functional of 2-dimensional Yang-Mills
theory, except for the parametric $x_1, x_2$- dependence of the fields.
Eq.  (\ref{S11}) can be regarded as the strong coupling limit
of Yang-Mills theory. This interpretation is consistent with the result
of ref. \cite{mansfield} where a strong coupling expansion of Yang-Mills
theory was performed and the leading order was found to be given by $D =
2$ Yang-Mills theory. This result is also confirmed in the field
strength approach \cite{RFSA}.
\bi

\no
It is now straightforward to evaluate in the reduced, 2-dimensional
Yang-Mills theory (\ref{S11}) a Wilson loop in the 0-3 plane. One finds
the area law in agreement with the linear rising potential between
static charges as found above.
\bi

\section{Elimination of the electric fields}
\bi 

\no
In the gauged fixed Yang-Mills theory, where the Gau\ss ' law
constraint has been fully resolved, the electric field variables occur
still only quadratically in the exponent, so that these variables can be
integrated out. The integral over $e_3$ is trivial. To perform the
integral over $E'_\perp$ it is convenient to introduce a more compact
notation. We define the kernel\footnote{Note that, up to an irrelevant
constant, $Det \hk K$ gives the square of the Faddeev-Popov 
determinant (\ref{D1a}) .} 
\be
\label{104}
K^{ab} (x, y) 
 = \lk
\begin{array}{cc}
K^{a_0 b_0} & 0 \\
0 & K^{\bar{a} \bar{b}}
\end{array}
\rk
= \lk
\begin{array}{cc}
- \partial'^{2}_{3} \delta^{a_0 b_0} & 0 \\
0 & - \lk \hat{d}_3 \hat{d}_3 \rk^{\bar{a} \bar{b}}
\end{array}
\rk
\delta^{(4)} (x - y) 
\hk . 
\ee
Furthermore, we define
\be
\lk 
E'_\perp (x) + e_\perp (\bar{x}) \rk^c = \int d^4 y P^{c c'} (x, y)
E^{c'}_{\perp} (y) + e^{(0) c}_{\perp} \hk ,
\ee
where $e^{(0)}_{\perp}$ is defined by eq. (\ref{106}).
and
\be
P^{c c'}_{\underline{i} \underline{j}} = \delta^{c c'} \delta_{\underline{i}
\underline{j}} \delta^{(4)} (x, y) - \delta^{c c_0} \lk
\nabla_{\underline{i}} \frac{1}{\triangle'_\perp} \rk_x \frac{1}{L} \int d x_3
\hat{A}'^{c_0 \bar{b}}_{\underline{j}} (x) \delta^{\bar{b} c'}  
\delta^{(4)} (x
- y) \hk .
\ee
This quantity fulfills the relations
\be
\nabla'_{\underline{i}} P^{\bar{a} b}_{\underline{i} \underline{j}} (x,
y) & = & \nabla'_{\underline{j}} \delta^{\bar{a} b} \delta^{(4)} (x -
y)\nonumber\\
\nabla'_{\underline{i}} P^{a_0 b}_{\underline{i} \underline{j}} (x,
y) & = & \frac{1}{L} \int  d x_3 \hat{D}'^{a_0 b_0}_{\underline{j}} (x)
\delta^{(4)} (x, y) \hk .
\ee
In this notation we have
\be
\label{G106}
&& \int d^4 x \left[ \lk \frac{1}{\partial'_3} \Gamma^{n}_{\perp} \rk^2
+ \lk \frac{1}{\hat{d}_3} \Gamma^{ch}_{\perp} \rk^2 \right] = \int d^4 x
\left[ \Gamma_\perp K^{- 1} \Gamma_\perp \right] 
 \nonumber\\
& = & \int d^4 x  \left[ \lk \hat{D}'_\perp \lk E'_\perp + e_\perp \rk +
\rho \rk K^{-1}  \lk \hat{D}'_\perp \lk E'_\perp + e_\perp \rk +
\rho \rk \right] \nonumber\\
& = & \int d^4 x \lk \hat{D}'_\perp \lk P E'_\perp + e^{(0)}_{\perp} \rk
+ J_0 \rk  K^{-1}  \lk
\hat{D}'_\perp \lk P E'_\perp  + e^{(0)}_{\perp} \rk + J_0 \rk\nonumber\\
& = & \int d^4 x \lk \lk E'_\perp P^T  + e^{(0)}_{\perp} \rk 
\lk - \hat{D}'_\perp \rk + J_0
 \rk
K^{-1} \lk \hat{D}'_\perp \lk P E'_\perp  + e^{(0)}_{\perp} \rk 
+ J_0 \rk \hk . 
\ee
Here, we have also introduced the transposed projector $P^T$ by 
\be
\lk P E'_\perp \rk^c (x) = \lk E'_\perp P^T \rk^c (x) \hk .
\ee
In this notation we can also write
\be
\int \lk E'_\perp E'_\perp + e^{2}_{\perp} \rk & = & \int \lk E'_\perp + 
e_\perp \rk^2 = \int \lk P E'_\perp  + e^{(0)}_{\perp} \rk^2 \nonumber\\
& = & \int \lk E'_\perp  P^T P E'_\perp + 2 e^{(0)}_{\perp} P E'_\perp + \lk
e^{(0)}_{\perp} \rk^2 \rk \hk .
\ee
In eq. (\ref{G106}) and below it is understood, that in $\hat{D}'_\perp
= \nabla_\perp + \hat{A}'_\perp$ the 2-dimensional gradient operator
$\nabla_\perp$ is replaced by the corresponding operator $\nabla'_\perp$
with the zero mode excluded
\be
\langle x_\perp | \vec{\nabla}'_\perp | x' \rangle = i
\sli_{\vec{n}_\perp \neq 0} \vec{n}_\perp e^{i \frac{2 \pi}{L}
\vec{n}_\perp \lk \vec{x} - \vec{x'} \rk} \hs , \hs \vec{n}_\perp =
(n_1, n_2) \hk .
\ee
This is admissible since $\nabla_\perp
E_\perp = \nabla'_\perp E_\perp$.
The integral over $E'_\perp$ can then easily be carried out yielding for
the transition amplitude\footnote{Note in eq. (\ref{G34}) from
$\hat{D}'_\perp e_{\perp}^{(0)} =
\nabla_\perp e^{(0)}_{\perp} + \hat{A}'_\perp e^{(0)}_{\perp}$ the term
$\nabla_\perp e^{(0)}_{\perp}$ can be dropped since this quantity does
not depend on $x_3$ and hence vanishes when acted on with $K^{- 1} \sim
\lk \partial'^{2}_{3} \rk^{-1}$}

\newpage
\be
\label{G34}
Z [J]
 & = & \int \cD \lk A'_\perp, a_3 \rk \pli_{x_0} Det^{ - \ehss}
H  \exp \left\{ \frac{i}{g^2} \left[ L \frac{1}{2} \int d^3 \bar{x}
\lk \partial_0
a_3 (\bar{x}) \rk^2 -  L
\int d^3 \bar{x} a_3 (\bar{x}) j_3 (\bar{x}) \right. \right. \nonumber\\
& & + \frac{1}{2} \int d^4 x \lk e^{(0)}_{\perp} - \partial_0 A'_\perp + 
\lk D'_\perp e^{(0)}_{\perp} + J_0 \rk K^{-1} D_\perp
 \rk P H^{- 1} 
 P^T \lk
e^{(0)}_{\perp} - \partial_0 A'_\perp \right. \nonumber\\
&  & - \left. D'_\perp K^{-1} \lk D_\perp
e^{(0)}_{\perp} + J_0 \rk 
\rk\nonumber\\
& & + \int d^4
x \lk e^{(0)}_{\perp} \partial_0 A'_\perp - \frac{1}{2} \lk
e^{(0)}_{\perp} \rk^2 \rk 
- \frac{1}{2} \int d^4 x \lk \hat{D'}_\perp
e^{(0)}_{\perp} + J_0 \rk \frac{1}{K} \lk \hat{D'}_\perp
e^{(0)}_{\perp} + J_0 \rk \nonumber\\
& & \left. \left. -  \frac{1}{2} \int d^4 x \lk B (A') \rk^2
\right]
\right\} \hk . 
\ee
Here, we have introduced the kernel
\be
\label{G35}
H = P^T P + P^T \lk - \hat{D}'_\perp \frac{1}{K} \hat{D}'_\perp \rk P 
\hk , 
\ee
which is defined in the space of the spatially periodic 2-dimensional
vector functions. 
\bi

\no
It is instructive to consider the limiting case $A'_\perp = 0$, where
the kernel (\ref{G35}) reduces to
\be
H = 1 - \nabla_\perp K^{-1} \nabla_\perp \hk .
\ee
In this case the terms of the action (\ref{G34}) containing the static source
$J_0$ reduce to 
\be
\left[ J_0 \frac{1}{K} \hat{D}'_\perp P H^{- 1} P^T \lk - 
\hat{D}'_\perp \rk  \frac{1}{K} J_0 - J_0  \frac{1}{K} J_0
\right]_{A'_\perp = 0}  = J_0 \frac{-1}{K - \triangle'_\perp} J_0
\ee
and in the limit $a_3 = 0$ we recover the familiar Coulomb law 
\be
\left[ J_0 \frac{-1}{K - \triangle'_\perp} J_0 \right]_{a_3 = 0} = J_0
\frac{1}{\triangle'} J_0 \hs , \hs \triangle' = \triangle'_\perp +
\partial^{2}_{3} \hk .
\ee
Note that in the continuum limit the zero eigenvalue of
$\triangle'$ disappears and $\triangle'\to \triangle$.
In the Yang-Mills theory the infrared singular behaviour of the static
Coulomb law is avoided by the presence of the abelian field $a_3
(\bar{x} )$. The infrared behaviour of Yang-Mills theory, and in
particular the confinement mechanism, should therefore be essentially
determined by this fluctuating field. This is in agreement with
the findings of the previous section.
\bi

\no
In Appendix B it is shown that for $A'_\perp = 0$
\be
Det^{- \ehti} H = 
Det^{\ehti} K \cdot Det^{- \ehti} \lk K - \Delta'_\perp \rk \hk .
\ee
Here
\be
Det^{\ehti} K = {\rm const.} Det^{\ehti} \lk - \hat{d}_3 \hat{d}_3 \rk 
= {\rm const} \cdot
J
\ee
is precisely the Faddeev-Popov determinant (\ref{D1a}) which, as shown
in Appendix A, up to an irrelevant constant coincides the Haar 
measure of $SU (N)$
\be
J = 
\pli_{k > l} \frac{1}{L^2} \sin^2 L \frac{\alpha_k (\bar{x}) -
\alpha_l (\bar{x})}{2} \hk ,\hk \sli_{k=1}^N \alpha_k(\bar x)
=0 \hk ,
\ee
where $\alpha_k (\bar{x})$ are the diagonal elements of $i \hk a_3
(\bar{x})$ (c.f. also ref. \cite{rein}).
\bi

\no
Note, that after imposing Gau\ss ' law we have obtained here the
Haar measure of the gauge group for the functional integral over the
gauge field $a_3 (\bar{x})$ although we started from the usual functional
integral representation with flat integration measure but gauge fixed by
the Faddeev-Popov method. This result is a manifestation of the
observation \cite{rein} that the standard functional integral
repesentation of Yang-Mills theory with gauge fixed by the Faddeev-Popov
method fully respects gauge invariance even in the non-perturbative
regime and that, in particular, the
Haar measure of the gauge group (necessary for a projection onto gauge
invariant states) naturally arises from the Faddeev-Popov determinant.
\bi 

\no
The Haar measure and hence $Det^{- \ehti} H$ vanish for degenerate field
configurations $a_3 \lk \bar{x} \rk$, for which two diagonal elements
coincide, i.e. $\lk a_3 \lk \bar{x} \rk \rk_{kk} = \lk  a_3 \lk \bar{x}
\rk \rk_{ll} $ for $k \neq l$.
\bi

\no
Since
\be
Det^{- \ehti} H = \exp \lk - \frac{1}{2} Tr \log H \rk
\ee
we receive an additional contribution to the effective action of the
remaining (physical) degrees of freedom, which represents an action
barrier to keep the system out of the singular field configurations.
Such (energy) barriers have been also found in alternative
formulations of gauge theory in terms of gauge invariant variables
\cite{johnson2,bauer}.
\bi

\no
Finally, let us make a few comments concerning the relation of the
present approach with those interpreting confinement as a dual Meissner
effect arising from monopole condensation. In fact, the gauge defined by
eq. (\ref{G16}) and (\ref{G24}) is a
variant of maximal Abelian gauge \cite{thooft,schier}. In these gauges 
monopoles arises at
those singular points in configuration space where the gauge fixing is
not unique. The gauge (\ref{G16}) is not 
unique at those singular points $x = x_S$,
where the field $a_3 (x)$ is degenerate, i.e. two eigenvalues of $a_3
(x)$ coincide. It is straightforward to show \cite{thooft,schier} that near
these singular points the gauge transformation $\Omega \lk \vec{x} \rk$
necessary to fulfill the gauge  (\ref{G16}) is such that the Abelian part of
$\Omega \partial_{\underline{\mu}} \Omega^\dagger (x)$ developes a
``magnetic'' monopole in $\underline{\mu} = 0, 1, 2$ space.
Field configurations for which the gauge fixing is not unique give rise
to zeros of the Faddeev-Popov determinant.
In fact
in the present case the Faddeev-Popov determinant
\be
Det i D_3 \sim {J} \sim  \sqrt{Det \hk K} \hk 
\ee
vanishes at the singular points of degenerate $a_3 \lk \bar{x} \rk$
field configurations.
The field configurations of vanishing Faddeev-Popov determinant define
the Gribov Horizon, which in the present context is therefore built up
from monopoles.
\bi

\section{Concluding remarks}
\bi

\no
In this paper we have considered $D=3+1$ dimensional Yang-Mills
theory defined on a spatial torus. Using a variant of `t Hooft's
maximum Abelian gauge (see eqs. (\ref{G16}), (\ref{G24}),
(\ref{G26})) we have 
performed a complete resolution of  Gau\ss ' law. This has
resulted in a functional integral representation of Yang-Mills
theory which is entirely defined in terms of unconstrained,
gauge fixed variables. These are the spatial gauge fields $A'_i,
~i=1,2$ defined by eqs. (\ref{58}), (\ref{E65}), (\ref{G39}), the neutral
$x_3$-independent part of which is transverse in 1-2-plane (see
eq. (\ref{G26})). In addition an Abelian $x_3$-independent field
$a_3(\bar x)$, defined by (\ref{G34b}) arises, so that the total
number of degrees of freedom is that of $2(N^2-1)$ fields. This
is the correct number of unconstrained degrees of freedom of
massless Yang-Mills theory with gauge group $SU(N)$. In this
respect the present approach is more efficient that the
pioneering works of refs. \cite{goldstone,faddeev} where
twice as much integration variables remain, since in that case
the electric field variables cannot be integrated out in closed
form. 

\bi
\no
We have worked here in the Hamilton functional integral
formulation which obviously violates Lorentz covariance.
Furthermore, the adopted gauge also violates spatial $SO(3)$
invariance but preserves axial symmetry. In this respect the
present gauge is advantageous over the gauge used in ref.
\cite{palumbo} which violates also axial symmetry. Of course a
complete elimination of all gauge degrees of freedom without
violating any space-time symmetry would be preferable. This has
been partly achieved in refs. \cite{johnson2,bauer} for $SU(2)$
in $D=3,4$. This appraoch works in the canonical Hamilton
(operator) approach, which obviously violates Lorentz covariance
but preserves all spatial symmetries. Unfortunately there is no
direct way to extend this approach to higher gauge groups
$SU(N>2)$, although some attempts have been undertaken for
$SU(3)$. For $SU(2)$ in $D=3$ a complete covariant, gauge
invariant description has been achieved in the so-called field
strength approach \cite{halpern} at the expense of a doubling of
the degrees of freedom \cite{RFSA}. We consider, however, the
violation of a global symmetry, which can easily
be restored, a minor problem. Of course, the exact Green
functions preserve all 
space-time symmetries even in gauges which violate these
symmetries. 
\bi

\no
As was illustrated in sect. 6 the field $a_3(\bar
x)$ represents the dominant infrared degrees of freedom, which
in particular are responsible for the emergence of the area law.
As a first step one might include only this Abelian field for a
study of the infrared sector of QCD. In fact, due to our adopted
gauge this is in the spirit of the Abelian dominance
observed in lattice calculations performed in
maximum Abelian type of gauge \cite{mon}.

\acknowledgments
{This work was partly performed during a visit at MIT. The author
thanks the members of the Center for Theoretical Physics, in
particular K. Johnson, for many interesting discussions and also
for the hospitality extended to him. Furthermore he gratefully
acknowledges useful discussions with M. Ilgenfritz,
F. Lenz and M. Thies. The author
thanks also L. Gamberg for a careful reading of the manuscript.
This work was supported in parts by DFG 856/1-3 and  by funds
provided by the U.S., Department of Energy (D.O.E.) under cooperative
agreement \# DE-FC02-94ER40818.}

\appendix

\section{Evaluation of the Faddeev-Popov Determinant}
\bi

\noindent
Below, we evaluate the functional determinant of $i \hat{d}_3$. Consider
the eigenvalue equation
\be
i \hat{D'}^{ab}_{3} \varphi^{b}_{\nu} = \mu_\nu \varphi^a_\nu \hk ,
\ee
where $\varphi_\nu$ has to satisfy periodic boundary conditions in
$x_3$. Multiplying this equation with the generators  in the fundamental
representation $T^a$ and using 
\be
T^a \hat{D}^{ab}_{\mu} \varphi^b = \left[ D_\mu , \varphi \right] \hs ,
\hs \varphi = \varphi^{a} T^a
\ee
the eigenvalue equation becomes
\be
i \partial_3 \varphi^{(\nu)} + \left[ i a_3, \varphi^{(\nu)} \right] =
\mu_\nu  \varphi^{(\nu)} \hk .
\ee
This equation is easily solved, since $i a_3 = i a^{c_0}_{3} T^{c_0}$ is
a (traceless) hermitian diagonal matrix, with real elements $\alpha_k
\equiv \lk a_3 \rk_{kk}$ satisfying $\sli^{N}_{k = 1} \alpha_k = 0$.
Hence, the
eigenvalue equation reads explicitly
\be
i \partial_3  \varphi^{(\nu)}_{kl} + \lk \alpha_k - \alpha_l \rk
\varphi^{(\nu)}_{kl} = \mu_\nu  \varphi^{(\nu)}_{kl} \hk.
\ee
The periodic eigenfunctions are given by $(\nu = (c,n))$
\be
 \varphi^{(\nu)}_{kl} = \eta^{c}_{kl} e^{i \omega_n x_3} \hs , \hs
\omega_n = \frac{ 2 \pi n}{L} \hs , \hs n = 0, \pm 1, \pm 2 \cdots \hk ,
\ee
where the $ \eta^{c}_{kl}$ denote the vectors of the Weyl basis of $SU
(N)$, which with $c= (r, s), \hs r, s = 1, 2, \dots , N$ is defined by
\be
 \eta^{c}_{kl} = \frac{1}{\sqrt{2}} \delta_{kr} \delta_{ls} \hk .
\ee
The corresponding eigenvalues read 
\be
\mu_\nu = \mu_{c, n} = \omega_n + \alpha_r - \alpha_s \hs c = (r, s)
\hk .
\ee
For fixed $n$, there are $2  \lk {N\atop 2}
\rk$ eigenvectors $\varphi^{\nu = (r, s)}$ corresponding to the
off-diagonal elements $r \neq s$. The corresponding eigenvalues come in
pairs $\omega_n \pm | \alpha_r - \alpha_s|$. Since $tr \varphi^{(\nu)} =
0$ there are only $N - 1$ independent eigenvectors $\varphi^{\nu = (r,
s)}$ with $r = s$, being degenerate with eigenvalue $\omega_n$. 
The total number of independent eigenvalues (for fixed $N$) is of course
$ N^2 - 1$. 
\bi

\no
We therefore obtain for determinant under consideration
\be
Det \lk i \hat{D}'_3 \rk & = & \pli_\nu \mu_\nu = \pli_{c^0} 
\pli^{\infty}_{n = -
\infty} \mu_{c^0,n} \nonumber\\
& = & \left[ \pli_{c^0} \pli_{n \neq 0} \omega_n
\right]
\left[ \pli_{r \neq s} \pli^{\infty}_{n = - \infty} \lk
\omega_n + \alpha_r - \alpha_s \rk \right]\nonumber\\
& = & Det \lk i \partial'_3 \rk \cdot Det \lk \hat{d}_3 \rk \hk .
\ee
Note that by definition of $\hat{D'}_3$ the mode
$n = 0$ has to be excluded for $r = s$, which is indicated by the
prime. The expression in the first bracket yields an irrelevant
(diverging) constant, which can be absorbed into the renormalization of
the functional integral.
\bi

\no
Using
\be
\sin x = x \pli^{\infty}_{n = 1} \lk 1 - \lk \frac{x}{\pi n} \rk^2 \rk
\hk ,
\ee
the expression in the second bracket can be transformed to 
\be
Det i \hat{d}_3 & = & \pli_{r \neq s} \lk \alpha_r -
\alpha_s \rk \pli^{\infty}_{n = 1} \left[ \lk  \alpha_r -
\alpha_s \rk^2 - \omega^{2}_{n} \right] \nonumber\\
& = & {\rm const}' \pli_{r \neq s} \lk \alpha_r -
\alpha_s \rk \pli^{\infty}_{n = 1} \lk 1 - \lk \frac{ \alpha_r -
\alpha_s}{\omega_n} \rk^2 \rk \nonumber\\
& = & {\rm const} \left[ \pli_{r \neq s} \frac{2}{L} \sin L \frac{
\alpha_r -
\alpha_s}{2} \right]\nonumber\\
& = & {\rm const} \pli_{r > s} \left[  \frac{2}{L} \sin L \frac{
\alpha_r -
\alpha_s}{2} \right]^{2} \hk .
\ee
Thus up to an irrelevant constant this determinant agrees 
for $\sum_{r=1}^N \alpha_r=0$ with 
the Haar measure of the group $SU (N)$ 
\be
\label{155}
J (L \alpha) = \pli_{k > l} \sin^2 L \frac{\lk \alpha_k -
\alpha_l \rk}{2} \hk . 
\ee
\bi


\section{Evaluation of ${\bf Det
\hk H}$}
\bi

\noindent
In what follows, we work out the functional determinant $Det \hk H$ for
$A'_\perp = 0$. In this limit $H$ is block diagonal in color space, i.e.
it has no matrix elements between the neutral and charged color
space 
\be
H^{ab} = \lk
\begin{array}{cc}
H^{a_0 b_0} & 0 \\
0 & H^{\bar{a} \bar{b}}
\end{array} \rk \hk .
\ee
This is because the same is true for the matrix $K$ (\ref{104}). Therefore, the
functional determinant $Det \hk H$ factorizes as
\be
Det \hk H = Det^{(n)} H Det^{(ch)} H \hk .
\ee
Since, the color neutral part of $H$ is given by
\be
H^{a_0 b_0}_{\underline{i} \underline{j}} = \delta^{a_0 b_0} \lk
\delta_{\underline{i} \underline{j}} + \nabla'_{\underline{i}}
\frac{1}{\lk \partial'_3 \rk^2} \nabla'_{\underline{j}} \rk \hk .
\ee
$Det^{(n)} H$ is an irrelevant constant, which will be ignored in the
following.
\bi 

\no
For the evaluation of $Det^{(ch)} H$ we consider the corresponding
eigenvalue equation
\be
H^{\bar{a} \bar{b}}_{\underline{i} \underline{j}}
\phi^{b}_{\underline{j}} (x) = \mu \phi^{a}_{\underline{i}} (x) \hk ,
\ee
which with the explicit form of $H^{\bar{a} \bar{b}}$ reads
\be
\label{162}
- \nabla'_i \lk {\bar K}^{-1} \rk^{\bar{a} \bar{b}}
\nabla'_{\underline{j}} \phi^{b}_{\underline{j}} (x) = (\mu - 1)
\phi^{a}_{\underline{i}} (x) \hk . 
\ee
Here, the eigenfunctions have to satisfy periodic boundary conditions
and we have introduced the abbreviation $\bar{K} = - \hat{d}_3
\hat{d}_3$, which is the matrix $K$ (\ref{104}) in the charged subspace.
\bi

\no
The eigenfunctions $\phi^{b}_{\underline{i}} (x)$ represent 2-dim.
spatial vectors, which we can split in longitudinal and transverse parts
\be
\phi^{b}_{\underline{i}} (x) = \phi^{b}_{\underline{i}} (x)^T +
\phi^{b}_{\underline{i}} (x)^L \hk ,
\ee
satisfying
\be
\partial_{\underline{i}} \phi^{b}_{\underline{i}} (x)^T = 0 \hs , \hs
\partial_{\underline{i}} \phi^{b}_{\underline{i}} (x)^L =
\partial_{\underline{i}} \phi^{b}_{\underline{i}} (x) \hk .
\ee
Obviously any (spatially) transverse vector function
$ \phi^{b}_{\underline{i}} (x)^T$ gives rise to a zero eigenvalue of
$\nabla_{\underline{i}} {K}^{-1} \nabla_{\underline{j}}$ and hence
to an eigenvalue $\mu = 1$ of $H$.
\bi

\no
The longitudinal part $ \phi^{a}_{\underline{i}} (x)^L$ gives rise to a
non-trivial eigenvalue $\mu \neq 1$. For these eigenvalues eq. (\ref{162}) can
be simplified.
\bi

\no
Operating on equation  (\ref{162}) from the left with $\nabla'_i$ and defining
\be
\varphi^{\bar{a}} (x) = \nabla'_{\underline{i}} \phi^{a}_{\underline{i}}
(x) \hk ,
\ee
the eigenvalue equation becomes
\be
- \Delta'_\perp \lk {K}^{-1} \rk^{\bar{a} \bar{b}} \varphi^{\bar{b}}
= (\mu - 1) \hk \varphi^{\bar{a}}
\hk .
\ee
Therefore the non-trivial part of the determinant of H is given by
\be
Det H = Det \lk 1 - \Delta'_\perp {K}^{-1} \rk \hk .
\ee
Note that the kernel of the r.h.s is a matrix in color and functional
space but a scalar in ordinary space, contrary to $H$ which is also a
matrix in the 2-dim. Euclidean space spanned by the 
$x_{\underline{i} = 1,2}$-axis.
The missing dimension on the r.h.s. is due to eigenvalues $\mu = 1$ of
$H$. For later use it will be convenient to separate off ${K}^{-1}$
yielding 
\be
Det H = \frac{Det \lk \bar{K} - \Delta'_\perp \rk}{ Det \bar{K}} \hk .
\ee
Since $i \hat{d}_3$ is the hermitian operator $K = -  \hat{d}^{2}_{3}$
is positive semi-definite, while $\lk - \Delta'_\perp \rk$ is strictly
positive definite. Therefore, $\lk \bar{K} - \Delta'_\perp \rk$ is a
positive definite operator and consequently $Det^{-\frac{1}{2}} \lk \bar{K} -
\Delta'_\perp \rk$ is non-singular, even for field configurations $a_3
(\bar{x} )$ for which $\bar{K}$ has zero eigenvalues. These field
configurations do not contribute to the transition amplitude due to the
presence of $Det^{\frac{1}{2}}\hk \bar{K}$. Note, that this 
determinant agrees with the
Haar measure of $SU (N)$ given in Appendix A.
\bi

\section{Residual gauge invariance}
\bi

\no
The spatially periodic boundary conditions to the gauge fields 
$A_i (x)$ restrict the gauge transformations 
$A_i \to A^{\Omega}_i$ to those gauge functions $\Omega (x)$, which
satisfy the equation
\be
\left[ D_i, \tilde{\Omega}_k (x) \right] = 0 \hk ,
\ee
where
\be
\label{170}
\tilde{\Omega}_k (x) = \Omega^\dagger \lk x + L {\bf e}^k \rk \Omega (x) \hk
.
\ee
For simplicity we have set here $L_1=L_2=L_3=L$.
This equation, which has to be fulfilled for all $A_i (x)$ and all $k
=1,2,3$, is solved for
\be
\label{G3}
\tilde{\Omega}_k (x) = Z_{n_k} \hk , 
\ee
with
\be
Z_n = e^{- \phi_n} = e^{i \frac{2 \pi n}{N}} \hs , \hs n = 0, 1, 2,
\dots, N - 1
\ee
being an element of the center of the gauge group. The center of the
group is defined by the set of elements commuting with all elements of
the group. Here $\phi_n$ is an element of the Cartan algebra $\cH$.
For $SU (2)$ the center of the group is given by
\be
\phi_{n = 0} = 0 & , & Z_0 = 1 \nonumber\\
\phi_{n = 1} = - i \pi \tau_3 & , & Z_1 = - 1 \hk .
\ee
\bi

\no
Eqs. (\ref{170}), (\ref{G3}) imply
\be
\label{G6}
\Omega \lk x + {\bf e}^k L \rk = Z^{*}_{n} \Omega (x) \hk . 
\ee
\bi

\no
The gauge conditions chosen above (see eqs. (\ref{G16}), (\ref{G24}),
(\ref{G26}) ) do not yet fix the gauge completely.
There is a residual gauge symmetry left, which will be exhibited below.
\bi

\no
First we note, that all three gauge fixing conditions are left unchanged
under permutations of the color indices $k, l$ of the fundamental
representation $ T^{a}_{k l} $ for the gauge group $SU (N)$. Such
permutations of the basis color vectors are generated by those global gauge
transformations $S \in SU (N)$, which are N-dimensional matrix 
representations of the
symmetry group $S_N$ (group of permutations of $N$ elements). 
\bi

\no
These transformations form the Weyl group. For $SU (2)$ the Weyl group
consists of two elements
$S_0 = 1$ and $S_1 = - i \tau^1 = 2 T^1$, which correspond,
respectively, to the trivial permutation 
and to an exchange of the two color indices.
The non-trivial permutation in fact represents $a$ rotation in color
space around the $1$-axis through an angle $\pi$
\be
S_1 = - i \tau^1 = e^{- i \pi \frac{\tau^1}{2}} = e^{\pi T^1} \hk .
\ee
Note, that $S_1$ is not an element of the Cartan algebra.
These considerations can obviously be extended
to larger gauge groups $SU (N > 2)$. 
\bi

\no
Besides the above discussed global residual gauge symmetry the gauge
conditions (\ref{G16}), (\ref{G24}), (\ref{G26}) are also invariant
under the so-called displacement symmetry \cite{guralnik,lenz1}. Let
us show how this comes about. The gauge condition $A^{ch}_{3} = 0$
leaves residual gauge transformations of the form 
\be
\label{G17}
\Omega = U (x) S \hs , \hs U (x) = e^{- \omega (x) } \hs , \hs \omega 
(x) \in \cH \hk , 
\ee
where $S\in S_N$ and $U (x)$ is a gauge transformation in the
maximal abelian subgroup (invariant torus).
\bi 

\no
The second gauge condition $\partial_3 A^{n}_{3} = 0$ is left invariant by
gauge transformations of the form (\ref{G17}) provided that 
\be
\partial_3 \partial_3 \omega (x) = 0 \hk .
\ee
This equation is satisfied if $\omega (x)$ is of the form
\be
\omega (x) = \omega^{(0)} (\bar{x}) + \omega^{(1)}   (\bar{x}) x_3 \hs ;
\hs \omega^{(0)} (\bar{x}) , \omega^{(1)}   (\bar{x}) \in \cH \hk .
\ee
The quasiperiodic boundary condition to the gauge function (\ref{G6})
requires
\be
e^{- \omega^{(1)}   (\bar{x}) L} = Z^{*}_{n_3} = e^{{\phi_{n_3}}} \hk .
\ee
This equation has to be fulfilled for all $\bar{x}$ and is solved for
\be
\label{G21}
\omega^{(1)}   (\bar{x}) = \frac{1}{L} \lk 2 \pi i k_3 - \phi_{n_3} \rk
:= \alpha_3 
\ee
with $k_3 = {\rm dia} \lk k^{(1)}_3,  k^{(2)}_3, \dots  k^{(N)}_3 \rk$
being a N-dimensional traceless diagonal matrix,
$\sli^{N}_{i = 1}  k^{(i)}_3 = 0$,
with integer entries $k_3^{(i)}$. 
\bi

\no
Finally, the third gauge constraint restricts the residual gauge
transformations to such function satisfying 
\be
\nabla_\perp \nabla_\perp \lk \omega^{(0)}_{(\bar{x})} + \frac{1}{2} L
\omega^{(1)} (\bar{x}) \rk = 0 \hk ,
\ee
which in view of equation (\ref{G21}) reduces to 
\be
\nabla_\perp \nabla_\perp \omega^{(0)} (\bar{x}) = 0 \hk ,
\ee
which implies
\be
\omega^{(0)} (\bar{x}) = \beta (x^0) + \alpha_\perp (x^0) x_\perp \hs ,
\hs \beta, \alpha_\perp \in \cH \hk .
\ee
The quasiperiodic boundary condition (\ref{G6}) requires the
$\alpha_{\underline{i}} (x^0)$ to fulfill the relation 
\be
e^{- \alpha_{\underline{i}} L} = Z^{*}_{n_{\underline{i}}} =
e^{\phi_{n_{\underline{i}}}} \hk ,
\ee
which is satisfied for 
\be
\alpha_\perp = \frac{1}{L} \lk 2 \pi k_\perp i - \phi_{n_\perp} \rk  \hk
\ee 
with $k_\perp = {\rm dia} \lk k^{(1)}_{\perp},  k^{(2)}_{\perp}, \dots ,
k^{(N)}_{\perp} \rk$ 
being again a traceless diagonal matrix with integer entries
$k^{(i)}_{\perp}$, 
satisfying  $\sli^{N}_{i = 1}  k^{(i)}_{\perp} = 0$.
\bi

\no
The residual gauge symmetry left after the three gauge constraints have
been implemented is therefore given by
\be
\Omega (x) = S e^{- \lk \beta + \vec{\alpha} \vec{x} \rk} \hk .
\ee
Besides the discret symmetry $S$ generating the permutation of color
indices the resi\-dual gauge symmetry consists of a global abelian gauge
transformation $e^{-  \beta}$ and the so-called displacement symmetry
$e^{- \vec{\alpha} \vec{x}}$. 
\bi

\end{document}